\documentclass[preprintnumbers, floatfix,letterpaper,aps,prd,epsfig,nofootinbib,
twocolumn
]{revtex4-1}
\usepackage{bm,graphicx,dcolumn,epstopdf,epsf, latexsym,mathbbol, amssymb,amsmath,color,slashed, mathrsfs,mathcomp,simplewick}
\usepackage[center]{subfigure}
\usepackage{multirow}
\usepackage{makecell}
\usepackage[colorlinks,linkcolor=blue,citecolor=blue,urlcolor=blue]{hyperref}

\begin{document}
\allowdisplaybreaks
 \newcommand{\bq}{\begin{equation}} 
 \newcommand{\eq}{\end{equation}}
 \newcommand{\bqn}{\begin{eqnarray}}
 \newcommand{\eqn}{\end{eqnarray}}
 \newcommand{\nb}{\nonumber}
 \newcommand{\lb}{\label}
 \newcommand{\f}{\frac}
 \newcommand{\p}{\partial}
\newcommand{\PRL}{Phys. Rev. Lett.}
\newcommand{\PLB}{Phys. Lett. B}
\newcommand{\PRD}{Phys. Rev. D}
\newcommand{\CQG}{Class. Quantum Grav.}
\newcommand{\JCAP}{J. Cosmol. Astropart. Phys.}
\newcommand{\JHEP}{J. High. Energy. Phys.}
\title{An analytical approach to the field amplification and particle production by parametric resonance during inflation and reheating}

\author{Tao Zhu${}^{a}$}
\email{zhut05@zjut.edu.cn} 

\author{Qiang Wu${}^{a}$}
\email{wuq@zjut.edu.cn; Corresponding author} 

\author{Anzhong Wang${}^{a, b}$}
\email{anzhong$\_$wang@baylor.edu} 

\affiliation{${}^{a}$ Institute for theoretical physics and Cosmology, Zhejiang University of Technology, Hangzhou, 310032, China
\\${}^{b}$ GCAP-CASPER, Physics Department, Baylor University, Waco, TX 76798-7316, USA}

\date{\today}

\begin{abstract}

Field amplification and particle production due to parametric resonance are highly nontrivial predictions of quantum fields that couple to an oscillating source during inflation and reheating. Understanding this two effects is crucial for the connection between the resonance phenomenon and precise observational data. In this paper, we give a general and analytic analysis of parametric resonance of relevant field modes evolving during inflation and reheating by using the uniform asymptotic approximation. This analysis can provide a clear and quantitative explanation for the field amplification and particle production during the resonance. The potential applications of our results to several examples, including sound resonance during inflation, particle productions during reheating, and parametric resonance due to self-resonance potentials, have also been explored. The formalism developed in this paper is also applicable to parametric resonance in a broad areas of modern science.

\end{abstract}
\maketitle

\section{Introduction} 

Parametric resonance is a resonance phenomenon that arises because some parameters of the system are varying periodically. In cosmology, it can occur in many scenarios during inflation and post-inflationary evolution. During inflation, the parametric resonance of inflationary perturbations can be induced by an effective oscillating sound speed which provides a mechanism for producing primordial black holes \cite{cai_primordial_2018b}. Similar resonance can also be trigged by an excited heavy field and produces features in the primordial spectrum and bispectrum \cite{gao_oscillatory_2013, saito_resonant_2012, saito_localized_2013a, chen_primordial_2012, chen_searching_2012}.  After inflation, the oscillating inflaton field around its potential minimum can lead to resonance in fields coupled to it, giving rise to copious particle production in various fields including standard model particles \cite{kofman_theory_1997, greene_structure_1997, bassett_geometric_1998}, primordial magnetic fields \cite{bassett_geometric_1998}, and gravitational waves \cite{easther_gravitational_2007}. In addition, a self-resonance potential can produce resonance in perturbations of inflaton itself and generate gravitational waves that could be within the forthcoming detections \cite{liu_gravitational_2018, antusch_oscillons_2018, fu_production_2018, zhou_gravitational_2013}. These complex and rich phenomena can be directly connected to precise observations, and a quantitative description and explanation of these resonance effects is a key ingredient for establishing it.
 
Normally, the important nontrivial effects due to parametric resonance are the amplification of the associated field modes and the corresponding particle productions at certain frequency ranges. Existing approaches for studying these two effects are either limited to numerical simulations or semi-analytic but qualitative which can be only applied to modes at resonance frequencies with certain conditions. One of important analytical approaches, in which the resonance is formulated in terms of successive scattering on the parabolic potentials, has been extensively adopted in the studying of the resonance phenomenon during reheating \cite{kofman_theory_1997, charters_phase_2005}. However, the treatment of this approach is based on the expansion of the parabolic like potential about its extreme points in each oscillation, so in principle it is only applicable to the broad type resonance \cite{kofman_theory_1997, charters_phase_2005}.  Our purpose in this paper is to present a quantitative and general analysis of parametric resonance of relevant field modes evolving during inflation and reheating by applying the uniform asymptotic approximation, an approximation that has been verified to be powerful and robustness in calculating primordial spectra for various inflation models \cite{zhu_constructing_2014, zhu_inflationary_2014, zhu_quantum_2014, zhu_power_2014, martin_kinflationary_2013,ringeval_diracborninfeld_2010, wu_primordial_2017, qiao_inflationary_2018, zhu_scalar_2015, zhu_detecting_2015, zhu_inflationary_2016, habib_inflationary_2002, habib_inflationary_2005, habib_characterizing_2004, zhu_inflationary_2014, zhu_highorder_2016, geng_schwinger_2018} and applications in quantum mechanics \cite{Zhu:2019bwj}. Our analysis does not only provide a quantitative derivation and explanation of the resonance conditions for the relevant modes, but also provide physical explanation for the field amplification and particle production rate which is valid at all frequencies. Our results can universally describe all the three types of the parametric resonance: the tachyonic, broad, and narrow resonances. This in turn provides a significant generalization of the analytical results obtained by using the successive parabolic scattering method in \cite{kofman_theory_1997, charters_phase_2005}, which is only valid for the broad resonance. Applications of our results to sound resonance during inflation, particle productions during reheating, and parametric resonance due to self-resonance potential have also been explored. Details of the formalism and calculations will be reported elsewhere.

\section{Uniform Asymptotic Approximation for Mathieu Equation}

\subsection{Approximate solution}

The evolution of associated field mode $u_k$ during inflation and reheating can be formally put in the form of the so-called Mathieu equation \cite{nationalinstituteofstandardsandtechnologyu.s._nist_2010} \footnote{The formalism developed in this paper can be easily extended to parametric resonance with other type equations, for example, to that with Lam\'{e} equation.}
\bqn\lb{mathieu}
\frac{d^2u_k}{dx^2}+ (A_k - 2 q \cos {2 x}) u_k=0.
\eqn
Here we consider the two parameters $A_k$ and $q$ as constant but as long as they are not varying too rapidly the assumption is reasonable. The field modes $u_k$ can has different meanings in different cosmological contexts and we will give its specific definition later in Sec. IV, when we discuss the parametric resonance with specific examples.

\begin{figure}
{\includegraphics[width=7.1cm]{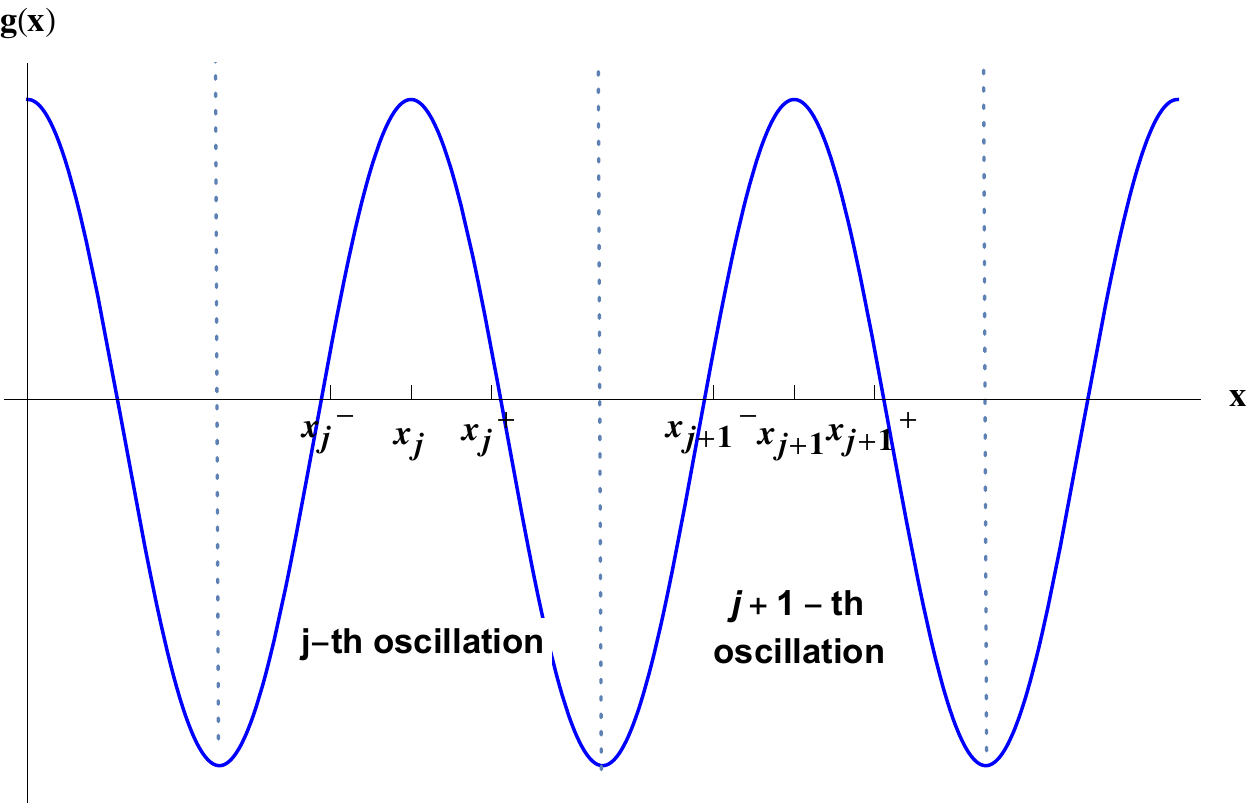}}
\caption{ A a schematic diagram for the $j$-th and $j+1$-th oscillations for the case with $A_k < 2 q$.}
 \label{gofx}
\end{figure}

In order to apply the uniform asymptotic approximation, let us write the above equation into the standard form \cite{olver_asymptotics_1997, zhu_inflationary_2014, olver_secondorder_1975}
\bqn\lb{eom_un}
\frac{d^2u_k}{dx^2} = \{g(x) + \mathfrak{q}(x)\}u_k
\eqn
with $g(x)+\mathfrak{q}(x) = 2q \cos{2x} - A_k$. Since the combination $g(x)+\mathfrak{q}(x)$ is regular, we can always choose $\mathfrak{q}(x)=0$ \cite{olver_asymptotics_1997, zhu_inflationary_2014, olver_secondorder_1975}. With this choice, we have
\bqn\lb{gx}
g(x)= 2q \cos{2x} - A_k,
\eqn
which is a periodic function, and in each oscillation, it has two turning points. Here we use $j$ to label the $j$-th oscillation, and the location of $\cos 2x=1$ is at $x_j$ during this oscillation. Then the corresponding two turning points (i.e. $g(x)=0$) are given by,
\bqn
x_{j}^{\pm}= \pm \frac{1}{2}\arccos{\frac{A_k}{2q}}+x_j.
\eqn
Obviously, the two turning points could be both real and single ($A_k <2 q$), double ($A_k=2q$ thus $x_{j}^{-}=x_{j}^{+}$), and complex conjugated ($A_k >2q$ thus $x_{j}^{-}= ( x_{j}^{+})^*$). We note that $x_j$ is the real part of $x_{j}^{\pm}$ when they are complexed conjugated. Fig.~\ref{gofx} provides a schematic diagram for the $j$-th and $j+1$-th oscillation for the case with $A_k < 2 q$ where the corresponding two turning points in each oscillation are both real.  

Since the function $g(x)$ is periodic, the general solution in each oscillation should take the same form. For this reason, we can only focus on the solution in the $j$-th oscillation, in which the function $g(x)$ has two turning points. Following \cite{zhu_inflationary_2014, olver_secondorder_1975},  we find that the general approximate solution of (\ref{eom_un}) in the $j$-th oscillation can be constructed in terms of parabolic cylinder function as
\bqn\lb{j_sol}
u_k^{j} &=& \left(\frac{\zeta_j^2-\zeta_0^2}{-g(x)}\right)^{\frac{1}{4}} \Big[a_j W\big(\frac{\zeta_0^2}{2}, \sqrt{2}\zeta_j\big)  +b_j W\big(\frac{\zeta_0^2}{2}, -\sqrt{2}\zeta_j\big) \big],\nb\\
\eqn
where 
\bqn
\zeta_0^2 &=&  \frac{2}{\pi} \int_{x_j^{-}}^{x^+_j} \sqrt{g(x)}dx=\frac{4}{\pi}\sqrt{2q-A_k} E(\mathcal{X}, \mathcal{Y}),\\
\mathcal{X} &\equiv& (i/2)\ln[(A_k- i \sqrt{4 q^2-A_k^2})/(2q)],\\
\mathcal{Y} &\equiv & 4 q /(2q-A_k)],
\eqn
with $E(\mathcal{X}, \mathcal{Y})$ denoting the incomplete elliptic integral of the second kind. 
The relation between variables $\zeta_j(x)$ and $x$ during the $j$-th oscillation is chosen such that $\zeta_j^2 -\zeta_0^2$ has the same types of zeros at the same places as those of $g(y)$. In each oscillation, the two zeros $\zeta_j(x_j^\pm)= \pm \zeta_0$ of the variable $\zeta_j$ exactly correspond to $x^{\pm}_j$. In this way,  $(\zeta_j^2 - \zeta_0^2 )/g(x)$ is ensured to be a regular function in each oscillation and the relation between $\zeta_j$ and $x$ can be determined via \cite{zhu_inflationary_2014, olver_secondorder_1975}
\bqn
\int_{{\rm Re } (x^{\pm}_{j+1})}^ x \sqrt{|g(x')|}dx' = \int_{\pm {\rm Re}(\zeta_0)}^\zeta \sqrt{|\zeta'^2 -\zeta_0^2|} d\zeta'.
\eqn
Obviously the sign of $\zeta_0^2$ is also sensitive to the nature of the turning points $x_j^{-}$ and $x_j^+$. $\zeta_0^2$ is positive when $x_j^{-}$ and $x_j^+$ are both real, $\zeta_0^2=0$ when $x_j^{-}=x_j^+$, and $\zeta_0^2$ is negative when $x_j^{-}$ and $x_j^+$ are complex conjugated. 

The upper bounds of the errors of the approximate solution (\ref{j_sol}) in each oscillation can be estimated from the computation of the variation of the error control function \cite{zhu_inflationary_2014, olver_secondorder_1975},
\bqn
\mathscr{I}(\zeta_j) &=&  \int_{\pm \zeta_0}^{\zeta_j} \left[\frac{5 \zeta_0^2}{4|v^2-\zeta_0^2|^{5/2}} - \frac{3}{4|v^2-\zeta_0^2|^{3/2}} \right]dv \nb\\
&& - \int_{x_j^{\pm}}^x \left[ \frac{ \mathfrak{q}(\tilde x)}{g(\tilde x)} - \frac{5 g'^2(\tilde x)}{16 g^3(\tilde x)} + \frac{g''(\tilde x)}{4 g^2(\tilde x)}\right] \sqrt{|g(\tilde x)|}d\tilde x.\nb\\
\eqn
Based on the analysis of it, one in principle can improve the approximate solution (\ref{j_sol}) through two ways. One way is to consider the high order uniform asymptotic approximation by incorporating the effects of $\mathscr{I}(\zeta_j)$ into the approximate solution (\ref{j_sol}), and the other way is to minimize $\mathscr{I}(\zeta_j)$ by choosing $\mathfrak{q}(x)$ properly. We will consider both ways to improve the approximation in our future works.

\subsection{Validity of the adiabatic condition}

During oscillations, the non-adiabatic evolution can occur due to the presence of the turning points and the extrema in each oscillation. Before we study the particle production and field amplification due to the non-adiabatic evolution, we would like first to provide a qualitative analysis of the adiabaticity of the field mode $u_k$ by using WKB approximation.

In general, the solution of the field mode $u_k$ of the equation
\bqn\lb{sode}
\frac{d^2 u_k}{dx^2} + w^2(x) u_k =0,
\eqn
can be approximately given in terms of the WKB solutions
\bqn\lb{wkb}
u_k^{\rm wkb}(x) \simeq \frac{\alpha_k}{\sqrt{2 w}} e^{-i \int w dx} + \frac{\beta_k}{\sqrt{2 w}}e^{i \int wdx},
\eqn
if the adiabatic condition is satisfied. In order to seek the adiabatic condition, we substitute the WKB solution into (\ref{sode}) and find,
\bqn
\frac{d^2 u_k^{\rm wkb}}{dx^2} + w^2(x) \left(1- \frac{w''}{2w^3} +\frac{3 w'^2}{4 w^4}\right) u_k^{\rm wkb}=0.
\eqn
Therefore, for the WKB solution (\ref{wkb}) to be a good approximation to the field mode $u_k$, the adiabatic condition
\bqn\lb{acon}
\epsilon_0 = \left|\frac{3w'^2}{4 w^4} - \frac{w''}{2 w^3}\right| \ll 1
\eqn
has to be satisfied.

Note that for the Mathieu equation (\ref{mathieu}), we have
\bqn
w^2(x) \equiv A_k - 2 q \cos 2x = - g(x),
\eqn
and the function $\epsilon_0$ measures how well ($\epsilon_0 \ll 1$) or how bad ($\epsilon_0 >1$) the adiabaticity of the field modes $u_k$ is during the oscillations. Then the adiabatic condition $\epsilon_0 \ll 1$ can be expressed in terms of $g(x)$ as
 \bqn\lb{adia}
\epsilon_0=\left |\frac{g''}{4g^2} - \frac{5 g'^2}{16 g^3} \right| \ll 1,
 \eqn
 which obviously can be violated when $g \to 0$. Depending on the parameters $(A_k, q)$, another case that violates the adiabatic condition occurs in the region around the maximum point $x=x_j$ and the minimum point $x=x_j+\pi/2$ of $g(x)$ in each oscillation. Around the maximum point $x_j$, it is easy to check that
 \bqn \lb{maximun}
 \epsilon_0 \simeq \left|\frac{g''(x_j)}{4 g^2(x_j)}\right| = \frac{2q}{(A_k-2q)^2} \ll 1
 \eqn
 leads to 
 \bqn
\begin{cases}
A_k > \sqrt{2q}+2q, & \;\; 0<q\leq \frac{1}{2} ,\\
0 < A_k < 2q -\sqrt{2q} \;\; {\rm or} \;\; A_k > \sqrt{2q}+2q, & \;\; q>\frac{1}{2}.
\end{cases}\nb\\
 \eqn
 Similarly, around the minimum point $x_j+\frac{\pi}{2}$ of $g(x)$, we have
 \bqn
 \epsilon_0 \simeq \left|\frac{g''(x_j+\pi/2)}{4 g^2(x_j+\pi/2)}\right| = \frac{2q}{(A_k+2q)^2} \ll 1.
 \eqn
 Thus  in order to satisfy the adiabatic condition, the allowed parameter space is 
  \bqn \lb{minimum}
\begin{cases}
A_k > \sqrt{2q}-2q, & \;\; 0<q\leq \frac{1}{2} ,\\
A_k > 0, & \;\; q>\frac{1}{2}.
\end{cases}
 \eqn
 In Fig.~\ref{adiabatic} we plot the function $\epsilon_0$ for three different resonance types, the tachyonic resonance, broach resonance, and narrow resonance respectively \footnote{We will define these three types of resonances in Sec. III}. In the following, by using the allowed parameter space derived in the above, we would like to find out the adiabatic regions that can be used to define the number of particle productions for each cases.
 
 \subsubsection{$A_k < 2q$}
 
 For this case, $g(x)$ has two real turning points and around these points the adiabatic condition (\ref{adia}) is obviously violated ($\epsilon_0 \to \infty$). Around the maximum point $x_j$, because $w$ is purely imaginary such that the field mode $u_k$ is a combination of the growing and decaying solutions even if the adiabatic condition is satisfied. This implies $u_k$ around $x_j$ is purely classical and there is no concept of particle associated with it. Therefore, the only region (the green region in the top panel of Fig.~\ref{adiabatic} that could satisfy $(\ref{adia})$ and use (\ref{wkb}) to define particle number is around the minimum point $x_j + \pi /2$ of $g(x)$ provided that $(A_k, q)$ is in the allowed parameter space (\ref{minimum}) with $A_k <2q$. 
 
 \subsubsection{$A_k \gtrsim 2q$}
 
 For this case, $g(x)$ has two coalescing complex conjugated turning points. The real part of turning points are also very close to the maximum point $x_j$. Around $x_j$, it is easy to see from (\ref{maximun}) and the middle panel of Fig.~\ref{adiabatic} that the adiabatic condition is strongly violated. Around the minimum point $x_j+\pi/2$ of $g(y)$, similar to the case with $A_k <2q$, the adiabatic condition can be fulfilled provided that $(A_k, q)$ is in the allowed parameter space (\ref{minimum}) with $A_k  \gtrsim 2q$. Thus the particle number can only be  defined using (\ref{wkb}) in the region around $x_j+\pi/2$, as shown in the green region in the middle panel of Fig.~\ref{adiabatic}. It is interesting to note that the green region of variable $x$ in this case is larger than that in the case with $A_k <2q$. 

\subsubsection{$A_k \gg 2q$}

\begin{figure}
{\includegraphics[width=8.1cm]{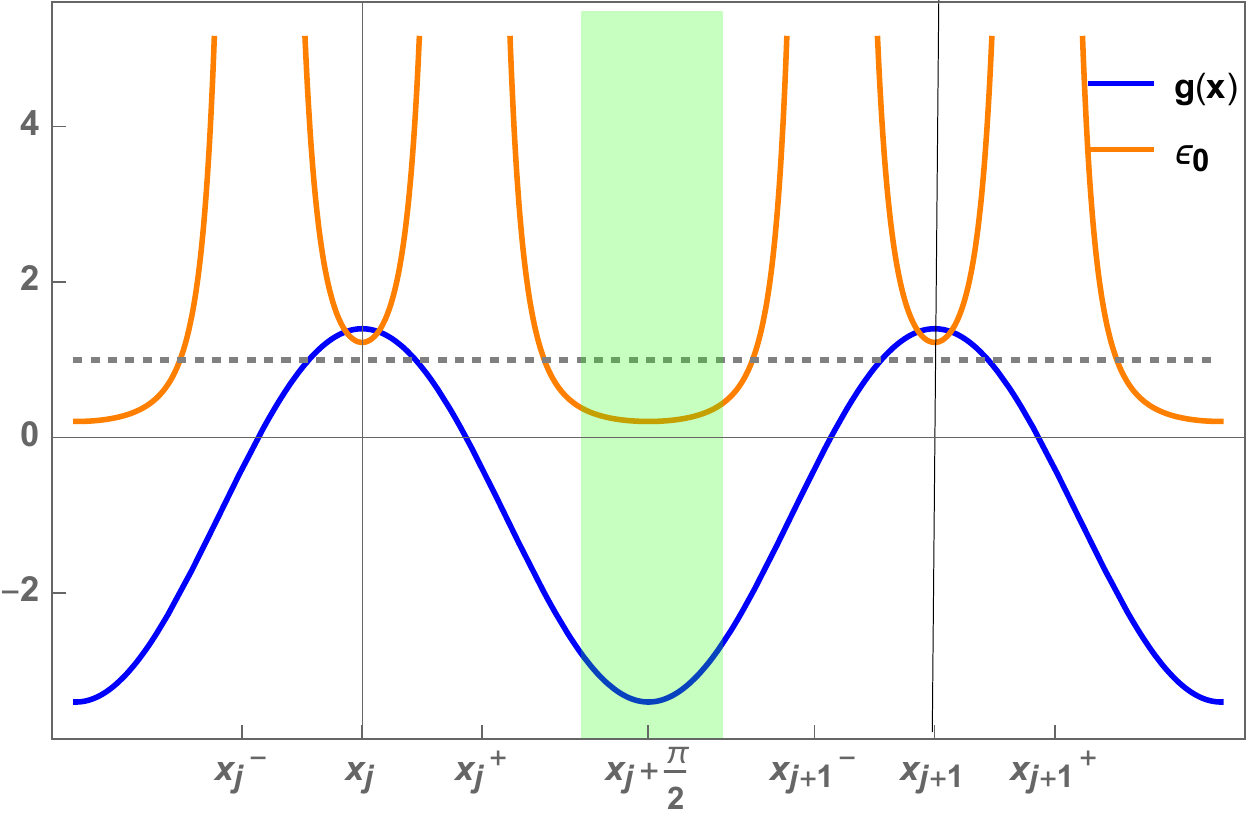}
\includegraphics[width=8.1cm]{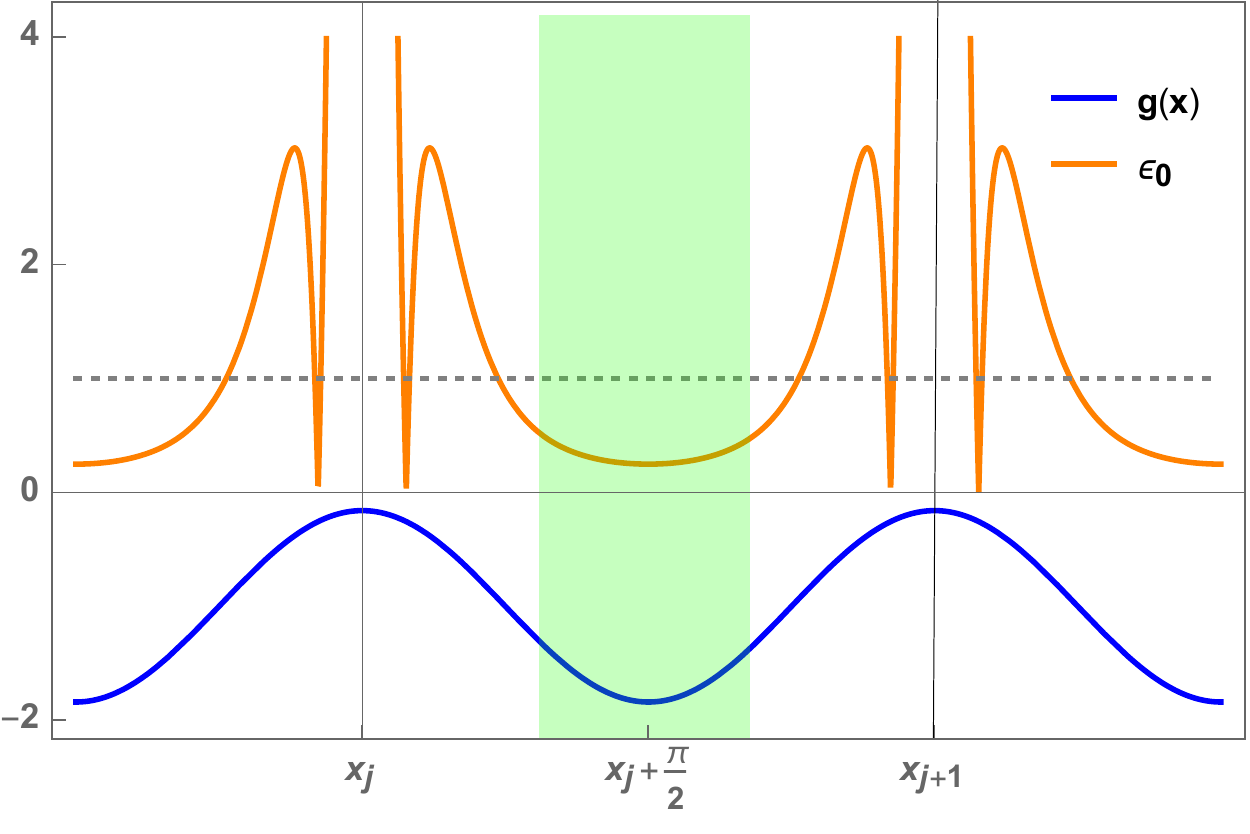}
\includegraphics[width=8.1cm]{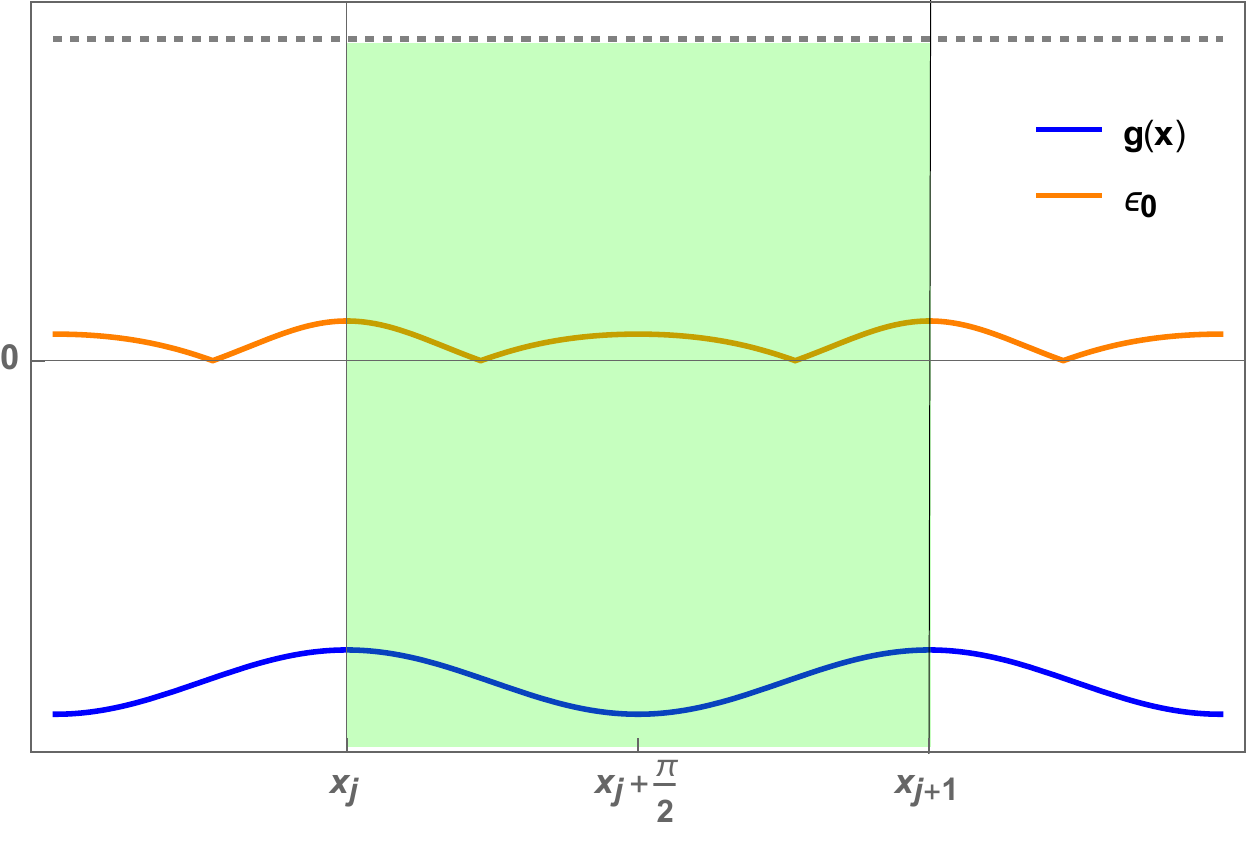}
}
\caption{ The quality function $\epsilon_0$ which measures the adiabaticity of the field modes for three types of resonance. The range of $x$ in the green region is the interval in each oscillation that the adiabatic condition $\epsilon_0 \ll 1$ is fulfilled and the number of particle is meaningful. Top panel: tachyonic resonance ($A_k<2q$) with $A_k=1$ and $q=1.2$. Middle panel: broad resonance ($A_k \gtrsim 2q$) with $A_k=1$ and $q=0.42$. Bottom panel: Narrow resonance ($A_k \gg 2q$) with $A_k=1$ and $q=0.05$. }
 \label{adiabatic}
\end{figure}

In this case, because $q \ll 1$, $\epsilon_0$ is always very small so that the adiabatic condition is completely fulfilled. Therefore, the particle number can be defined using (\ref{wkb}) during the whole region in each oscillation, as shown by the green region in the bottom panel of Fig.~\ref{adiabatic}. 

\begin{figure}
{\includegraphics[width=7.1cm]{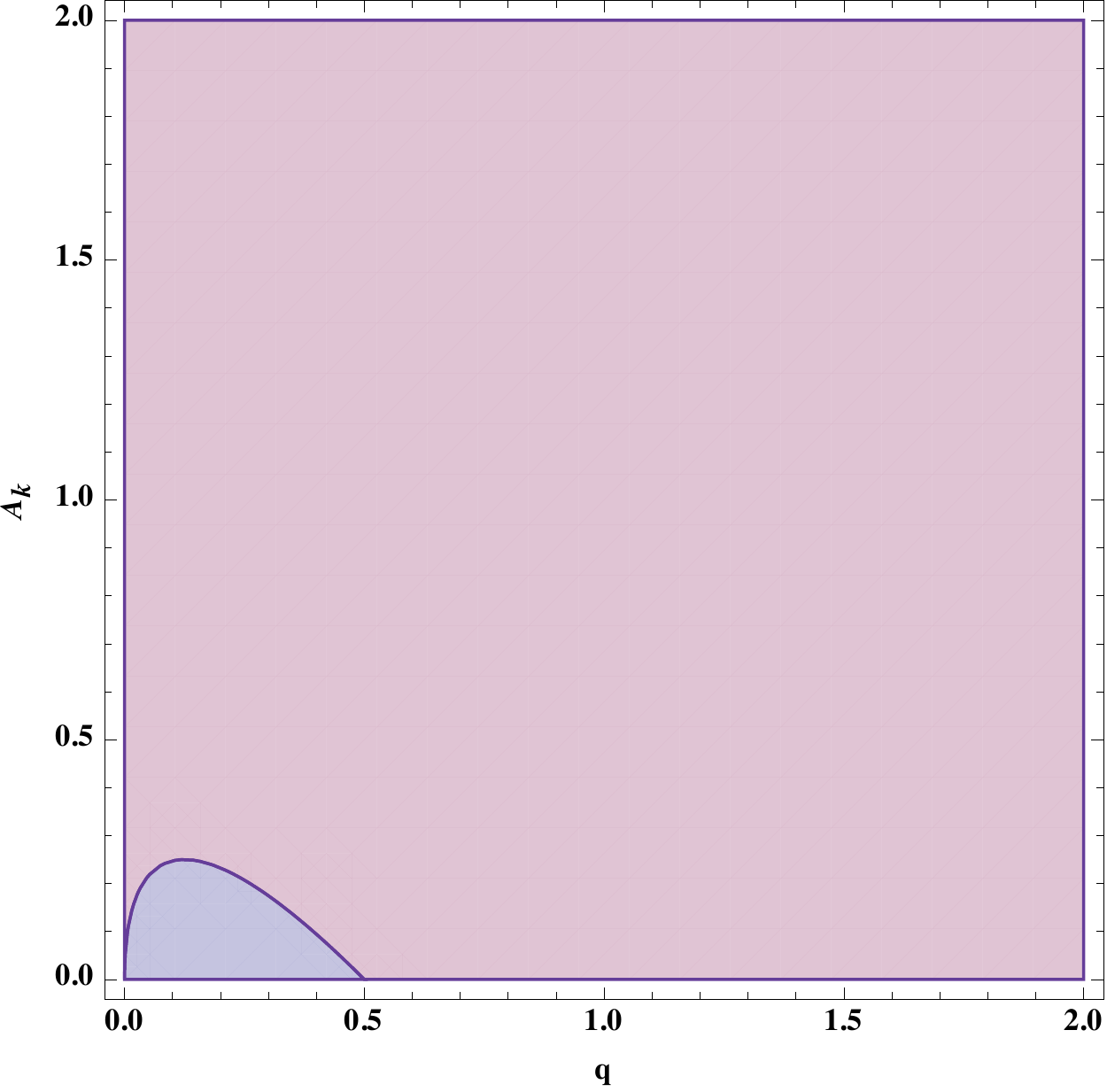}
}
\caption{The parameter space (the purple region) of $(A_k, q)$ that satisfies the adiabatic condition $\epsilon_0<1$ at the minimum point $x_j+\pi/2$.}
 \label{para}
\end{figure}

In summary, in order to use (\ref{wkb}) to compute the particle number associated with $u_k$, one has to impose two requirements: the adiabatic condition (\ref{adia}) and the real frequency $w$. According to the analysis in the above, the best place that can be used to define the particle notion for all the above three cases is the region around the minimum point $x_j$ with $(A_k, q)$ satisfies (\ref{minimum}). In Fig.~\ref{para}, we display the parameter space of $(a_k, q)$ (the purple region in the figure) that satisfies the adiabatic condition $\epsilon_0<1$ at the minimum point $x_j+\pi/2$. The small space of $(A_k, q)$ (the blue region in the figure) that does not satisfy the adiabatic condition corresponds to small values of $A_k$, which is out of the scope of the current paper because most of systems with parametric resonance in the reheating process are related to $A_k \geq 1$. In this paper, we only focus on the case with $A_k \geq 1$.

\subsection{Particle production rate and the field amplification}

In this subsection, we derive the general expression for both the particle production and the field amplification by using the approximate solution (\ref{j_sol}) of $u_k^j(x)$. Connecting the two nearby oscillating regions, namely connecting $u_k^j$ and $u_k^{j+1}$, gives the recursion relation between the coefficients $(a_{j+1}, b_{j+1})$ and $(a_j, b_j)$, 
\bqn
\begin{pmatrix}
a_{j+1}\\
b_{j+1} 
\end{pmatrix}
=\begin{pmatrix}
\kappa \sin{\mathfrak{B}}, & - \cos{\mathfrak{B}} \\
\cos{\mathfrak{B}}, & \kappa^{-1}\sin{\mathfrak{B}}
\end{pmatrix}
\begin{pmatrix}
a_{j}\\
b_{j} 
\end{pmatrix},\lb{recursion}
\eqn
where 
\bqn
\kappa = \sqrt{1+e^{\pi \zeta_0^2}} -e^{\pi \zeta_0^2/2},
\eqn
and
\bqn
\mathfrak{B} = 2 i \sqrt{2q-A_k} [E(\mathcal{X}, \mathcal{Y}) - E(\mathcal{Y})] + \frac{\pi}{2} + 2 \phi\left(\frac{\zeta_0^2}{2}\right),\nb\\
\eqn
when $x_{j}^{\pm}$ are both real, and
\bqn
\mathfrak{B} = 2 \sqrt{2q -A_k} E(\mathcal{Y})+\frac{\pi}{2}+2 \phi\left(\frac{\zeta_0^2}{2} \right),
\eqn
if $x_{j}^{\pm}$ are complex conjugated.

With above recursion relation, we are able to estimate the particle production rate during oscillations. A standard way for this purpose is to implement canonical quantization of the field mode $u_k(x)$, which satisfies the standard normalization condition $\langle u_k,u_k \rangle=i (u_k^* \dot u_k - \dot u^{*}_k u_k) =1$. When the adiabatic condition is fulfilled, one can expand the field mode in terms of adiabatic positive and negative frequency modes. In this way, if the initial state of the field mode is described by the adiabatic positive frequency mode, then the non-adiabatic evolution of the field mode due to the parametric resonance can lead to the generation of the excited state, which is a mixture of both the positive and negative frequency mode. The information of the particle production rate can be extracted from the projection of this excited state onto the adiabatic negative frequency modes \cite{geng_schwinger_2018}.

 As we have mentioned above, these non-adiabatic evolutions of the field mode leads to conspicuous particle production. This is also the main mechanism for the field amplification and particle production during the reheating. In order to calculate the particle production rate and field amplification during the parametric resonance, we first need to expand the field $u_k$ in terms of both the adiabatic positive and negative frequency modes in the adiabatic region.  Around the end of the $j+1$-th oscillation (i.e. $x=x_{j+1}+\pi/2$), as we shown in the above subsection, the adiabatic condition is fulfilled if (\ref{minimum}) is satisfied and we can expand the approximate solution of mode function $u_k(x)$ as
\bqn
u_k(x) &=& \frac{\alpha_{j+1}}{\sqrt{2} (-g)^{1/4}} e^{-i \int_{x_{j+1}+\frac{\pi}{2}}^x \sqrt{-g(x')}dx'} \nb\\
&&+ \frac{\beta_{j+1}}{\sqrt{2} (-g)^{1/4}} e^{i \int_{x_{j+1}+\frac{\pi}{2}}^x \sqrt{-g(x')}dx'},
\eqn
where $\alpha_{j+1}$ and $\beta_{j+1}$ are two Bogoliubov coefficients, which can be related to $a_{j+1}$ and $b_{j+1}$ by comparing both the uniform asymptotic approximate solution and WKB solution near the end of the $j+1$-oscillation. The modes $$ \bar u_k^{\pm}(x) = \frac{1}{\sqrt{2} (-g)^{1/4}} e^{ \mp i \int_{x_{j+1}+\pi/2}^x \sqrt{-g}dx'} $$ denote two basic adiabatic positive and negative frequency modes. With this procedure, the particle production rate $n_{k}^{j+1} $ can be calculated via
\bqn\lb{particle0}
n_k^{j+1} &=& \langle \bar u_k^-, u_k \rangle  \langle  u_k, \bar u_k^- \rangle = |\beta_{j+1}^2|,
\eqn
which leads to the following recursion relation,
\bqn\lb{particle}
n_k^{j+1} &=&e^{\pi \zeta_0^2}+(1+2 e^{\pi \zeta_0^2})n_k^{j}+2 e^{\pi \zeta_0^2/2} \sqrt{1+e^{\pi \zeta_0^2}} \nb\\
&&~ \times \sqrt{n_k^{j} (1+n_k^{j})} \cos{(2 \Theta +2 \mathfrak{B}-{\rm ph}\alpha_j+{\rm ph}\beta_j)},\nb\\
\eqn 
where $\Theta=-\mathfrak{B}/2$. This result is general and accurate, which can be also approximately reduced to those given in \cite{kofman_theory_1997, charters_phase_2005} with condition $\zeta_0^2 \lesssim 0$.

\begin{figure}
{\includegraphics[width=8.5cm]{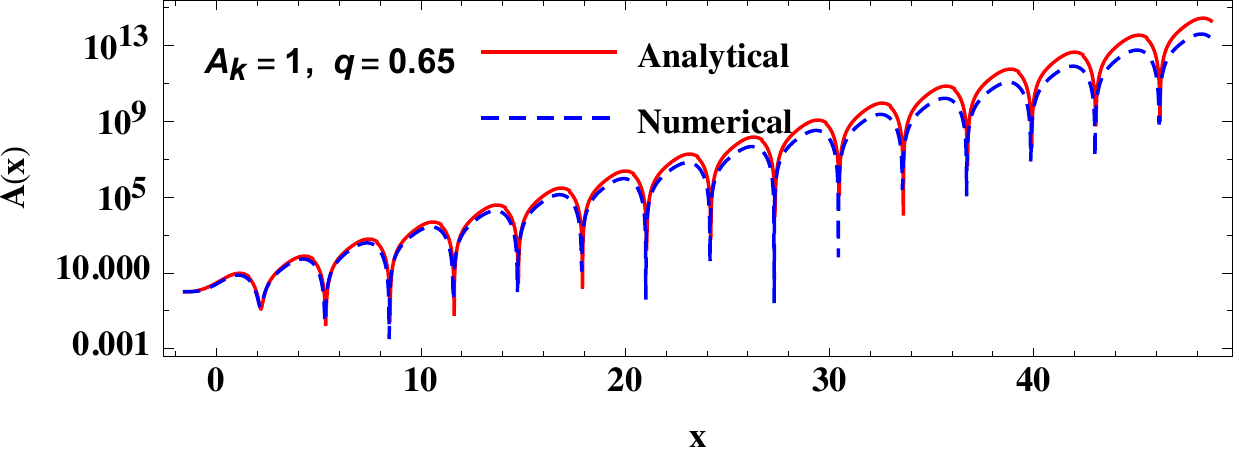}}
{\includegraphics[width=8.5cm]{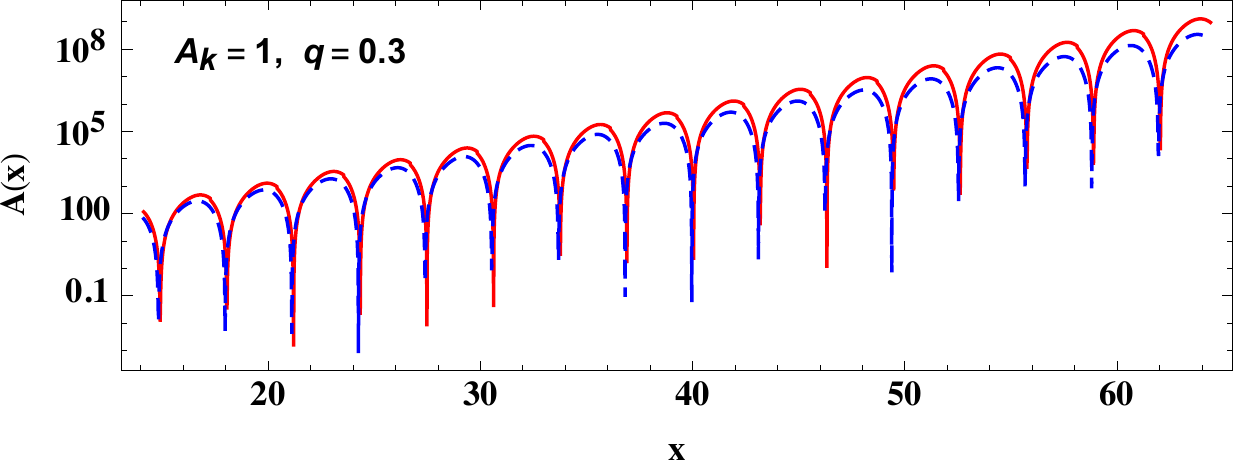}}
{\includegraphics[width=8.5cm]{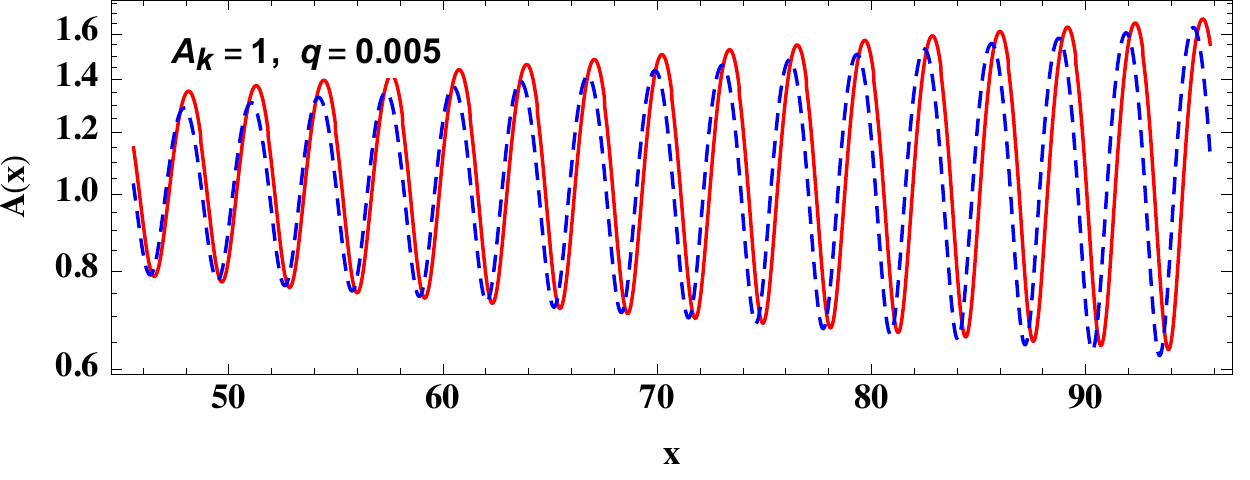}}
\caption{The analytic (red curve) and numerical (blue dashing curve) evolutions of the field amplification $A(x)=|u_k(x)/u_k(x_1-\pi/2)|^2$ at resonance bands during several number of oscillations. Top panel: tachyonic resonance during the first 10 oscillations. Middle panel: broad resonance during the 5-th to the 20-th oscialltions. Bottom panel: narrow resonance during the 15-th to 30-th oscillations.}
 \label{fit}
\end{figure}

Using the recursion relation (\ref{recursion}), one could relate the $(j+1)$-th oscillation to $(j+1-N)$-th oscillation via
\bqn\lb{connection}
a_{j+1} + Y_{\pm} b_{j+1} &=& Z_{\pm}^{N} (a_{j-N}+ Y_{\pm} b_{j-N} ),
\eqn
where 
\bqn
Y_{\pm } &=& -  \frac{\sec{\mathfrak{B}}}{2 \kappa}\Big [(\kappa^2-1) \sin{\mathfrak{B}} \nb\\
&&~~~~~~  \pm \sqrt{ (\kappa^2-1)^2\sin^2{\mathfrak{B}} - 4 \kappa^2 \cos^2{\mathfrak{B}}}\Big], \\
Z_{\pm} &=& -\frac{1}{2 \kappa} \big[(-1-\kappa^2) \sin{\mathfrak{B}}  \nb\\
&&~~~~~~ \pm \sqrt{ (\kappa^2-1)^2\sin^2{\mathfrak{B}} - 4 \kappa^2 \cos^2{\mathfrak{B}}}\big],
\eqn
 with $Y_+ Y_{-}=1=Z_{+} Z_{-} $. This relation helps us to determine the coefficients $a_{j+1}$ and $b_{j+1}$ of the approximate solution (\ref{j_sol}) from the initial state. Fig.~\ref{fit} shows the evolution fo the field amplification $A(x)\equiv |u_k(x)/u_k(x_1-\pi/2)|^2$ from analytical solution (\ref{j_sol}) with coefficients derived from (\ref{connection}) that fits numerical results very well. However, we have to mention that, the relation (\ref{connection}) is only valid when $N$ is not vey large ($N \lesssim 30$ as shown by Fig.~\ref{fit} for examples).  If $N$ is large enough, the small error of the approximation can accumulate in each oscillation, thus it will becomes large enough to destroy the validity of (\ref{connection}). Formally, from (\ref{connection}) we can still derive both the field amplification $A_{j+1}=A(x_{j+1}+\frac{\pi}{2})$ and particle production rate $n_k^{j+1}$ after $j+1$ oscillations,
\bqn
&& n_k^{j+1} =  \frac{[(Y_++\kappa^2 Y_-)Z_+^{j} - (Y_-+\kappa^2 Y_+)Z_-^{j}]^2}{4 (Y_{+}-Y_{-})^2 \kappa^2}, \lb{betaj1}\\
&&A_{j+1}= \frac{(Y_+^2 + \kappa^2)(\sin \Theta + \kappa Y_{-} \cos \Theta )^2}{\kappa^2 (Y_+ - Y_-)^2} Z_{+}^{2 j}  \nb\\
&&~~~~~~ +  \frac{(Y_-^2 + \kappa^2)(\sin \Theta + \kappa Y_{+} \cos \Theta )^2}{\kappa^2 (Y_+ - Y_-)^2} Z_{-}^{2 j} \nb\\
&&~~~~~~ - 2 \frac{(1+\kappa^2)(\sin \Theta + \kappa Y_{+} \cos \Theta )(\sin \Theta + \kappa Y_{-} \cos \Theta ) }{\kappa^2 (Y_+ - Y_-)^2}. \nb\\
\lb{Aj1}
\eqn
Here we assume the oscillations start at $x_1-\frac{\pi}{2}$ and the field mode $u_k$ is at the BD vacuum state at this point. As we mentioned, these formulas are accurate only when the number of oscillations $j+1$ is not very large. For large number of oscillations, these formulas will be not accurate enough but can show great insight qualitatively in the analysis of resonance as we shown later. 

\section{Conditions for parametric resonance}
\begin{figure}
{\includegraphics[width=8.5cm]{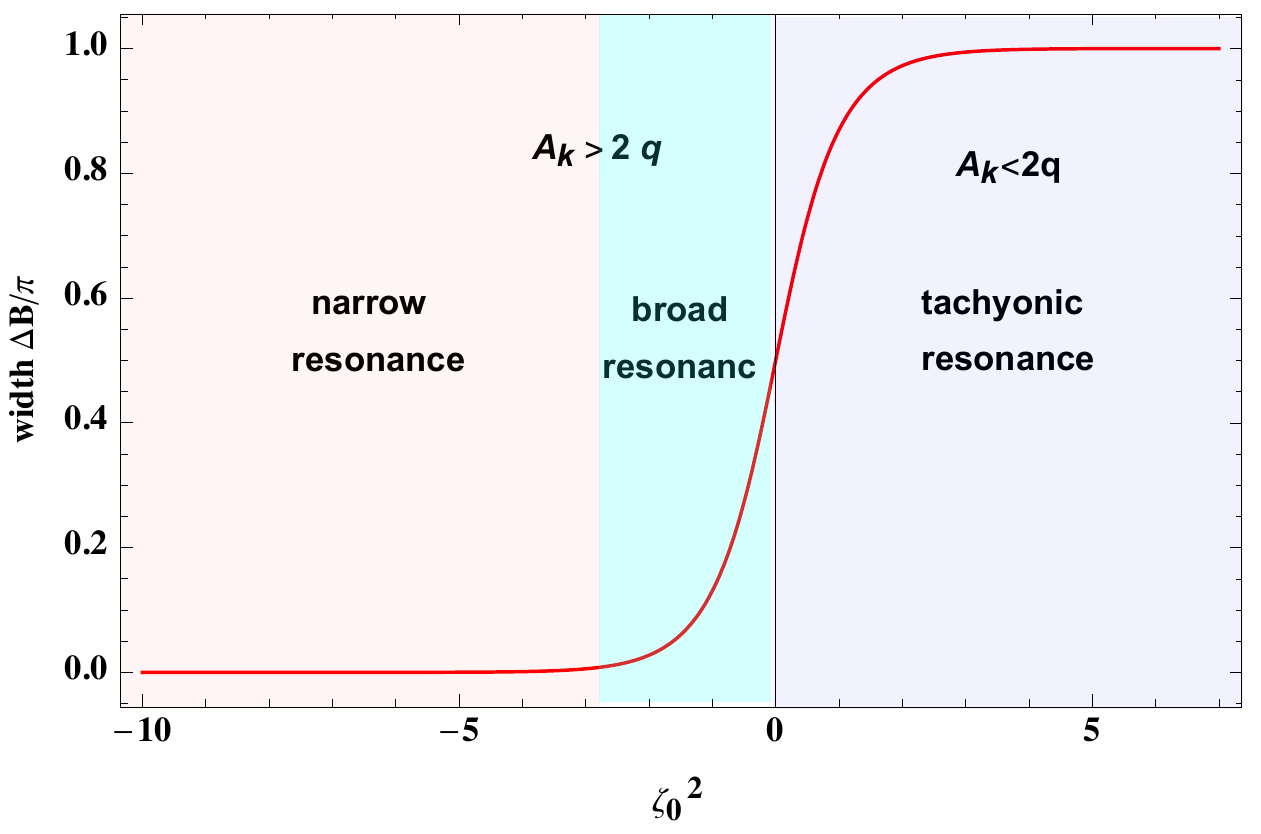}}
\caption{The fraction width $\Delta \mathfrak{B}/\pi$ in the region of parametric resonance as a function of $\zeta_0^2$.}
 \label{widthB}
\end{figure}

We are interest in amplification of the field mode $u_k$ during the parametric resonance. Normally, the strength or the amount of the field amplified during the process is very sensitive to the parameters $A_k$ and $q$. The essential conditions for the  resonance are those lead to the growing of modes in every oscillation. With this consideration, one has to require that
\bqn\lb{condition}
|Z_{+}|>1 \;\; {\rm or}\;\; |Z_{-}|>1.
\eqn
Since $Z_{+}Z_{-}=1$, we observe that $|Z_{\pm}|=1$ when $Z_{\pm}$ are complex conjugated. Thus for the field mode $u_k$ growing in every oscillation, $Z_{\pm}$ have to be both real, which in turn leads to the requirement of positivity of terms under the square root of $Z_{\pm}$, that is
\bqn \lb{consition}
\tan^2{\mathfrak{B}} > e^{- \pi \zeta_0^2}.
\eqn
This leads to
\bqn\lb{resonance_band}
n \pi +{\rm arctan}\big(e^{-\frac{\pi \zeta_0^2}{2}}\big) < \mathfrak{B}<n\pi + \pi - {\rm arctan}\big(e^{-\frac{\pi \zeta_0^2}{2}}\big),\nb
\eqn
where $n$ is an integer which defines a series of instability bands labeled by $n$. From this condition, we can derive the width of the $\mathfrak{B}$ in each instablity band, which is 
\bqn
\Delta \mathfrak{B} = \pi - 2 {\rm arctan}\big(e^{-\pi \zeta_0^2/2}\big).
\eqn 
This shows explicitly that the width $\Delta \mathfrak{B}$ in each band essentially depends on $\zeta_0^2$ as shown in Fig.~\ref{widthB}.

Depending on the value of $\zeta_0^2$, the instability bands can be divided into three different cases, the tachyonic resonance, the broad resonance, and the narrow resonance, as shown in Fig.~\ref{widthB}. The evolution of $A(x)$ with the three representative cases have been illustrated in Fig.~\ref{fit} by comparing analytic and numerical results.

\subsection{Tachyonic resonance}
The tachyonic resonance corresponds to the modes with $0<A_k < 2q$, thus $g(x)$ has two real turning points in each oscillation and $\zeta_0^2 >0$. These modes cross periodically the tachyonic region ($g(x)>0$) which leads to exponentially growing for most of modes during each oscillation (c.f. the top panel of Fig.~\ref{fit}).  It can provide an important mechanism for the physical constructions of the tachyonic preheating models after inflation, see \cite{greene_inflaton_1997, Kofman:2001rb, Abolhasani:2009nb, Dufaux:2006ee} and references therein. For modes with $e^{\pi \zeta_0^2} \gg 1$, we find
\bqn
n_k^{(j+1)} &\simeq& e^{(j+1) \pi \zeta_0^2} (4 \sin^2 \mathfrak{B})^j, \lb{nj1_tach}\\
A_{j+1} &\simeq& 4 e^{(j+1) \pi \zeta_0^2} (4 \sin^2 \mathfrak{B})^j \sin^2{\Theta}.\lb{Aj1_tach}
\eqn
We observe that the particle production and field amplification are both exponentially enhanced which arises from two effects: the value of $e^{\pi \zeta_0^2}$ and the number of oscillations. The corresponding amplification becomes dramatically enhanced as $e^{\pi \zeta_0^2}$ increases. 

The analytical analysis of the tachyonic resonance has also been performed in the studying of the the tachyonic preheating process after the inflation by using WKB approximation \cite{Abolhasani:2009nb, Dufaux:2006ee}. The particle production rate $n_k$ after $j+1$ oscillations of inflaton reads (Eq. (16) in \cite{Dufaux:2006ee})
\bqn
n_k^{(j+1)} &\simeq& e^{2 (j+1) X_k} (2 \cos \Theta_k)^{2j}
\eqn
with $X_k=\pi \zeta_0^2/2$ and $\Theta_k = \mathfrak{B}+\pi/2 - \phi(\zeta_0^2/2)$. Considering $\phi(\zeta_0^2/2) \ll 1$ for $e^{\pi \zeta_0^2} \gg 1$ \cite{nationalinstituteofstandardsandtechnologyu.s._nist_2010}, we find $\sin^2{\mathfrak{B}} \simeq \cos^2{\Theta_k}$. Therefore we observe that the expression (\ref{nj1_tach}) which derives from (\ref{betaj1}) recovers the results in \cite{Dufaux:2006ee} when $e^{\pi \zeta_0^2} \gg 1$ for  tachyonic preheating scenarios. These results have been shown to be very accurate \cite{Abolhasani:2009nb, Dufaux:2006ee} in comparison with numerical results for a lot of concrete tachyonic preheating models  and thus also provides another support for our approximation derived in this paper. It is also worth noting that, while the validity of the results derived in \cite{Abolhasani:2009nb, Dufaux:2006ee} are limited to the tachyonic resonance with $e^{\pi \zeta_0^2} \gg 1$, the general results in eqs.~ (\ref{betaj1}) and (\ref{Aj1}) are valid for all three types of resonances. 

\subsection{Broad resonance}

The broad resonance corresponds to the modes with $A_k \gtrsim 2q$, for which $g(x)$ has two coalescing complex conjugated turning points and $\zeta_0^2 \lesssim 0$. For this case, the growth of the modes is weaker than that in the tachyonic instability band (c.f. the middle panel of Fig.~\ref{fit}). However, the amplification of the field modes can still be very efficient after a lot of oscillations. In fact, the broad resonance is the essential building block for understanding the copious particle productions and field amplification during the reheating phase right after the inflation \cite{kofman_theory_1997, greene_structure_1997, charters_phase_2005}. For broad resonance, the particle production for field mode $u_k$ in the instability band is sufficient after a few number of oscillations. With this fact we have $n_k^{j} \gg 1$, thus we can express the recursion relation for particle production rate in (\ref{particle}) as
\bqn
n_k^{j+1} \simeq n_k^j e^{2\pi \mu_k^j},
\eqn
where $\mu_k^j$ is the growth index and is given by
\bqn
\mu_k^j \simeq \frac{1}{2\pi} \ln \Big[1+2e^{\pi \zeta_0^2} + 2 e^{\pi \zeta_0^2/2} \sqrt{1+e^{\pi \zeta_0^2}} \nb\\
\times \cos (2 \Theta +2 \mathfrak{B}-{\rm ph}\alpha_j+{\rm ph}\beta_j) \Big]. \lb{grow}
\eqn    
This result is similar to that obtained in \cite{kofman_theory_1997, charters_phase_2005} (c.f. Eq. (16) in \cite{charters_phase_2005} and (58) in \cite{kofman_theory_1997}) by applying the theory of parabolic scattering but in fact they are not exactly the same. This is because the results in \cite{kofman_theory_1997} can be only considered as a specific case of our results by expanding (\ref{gx}) about the $x_j$ as 
\bqn
g(x) \simeq g_0 + \frac{1}{2}g_2 (x-x_j)^2. \lb{expansion}
\eqn
With this expansion and dropping $\phi(\zeta_0^2/2)$ terms in the expressions of $\mathfrak{B}$ and $\Theta$ since $\phi(\zeta_0^2/2) \ll 1$, it is easy to show that (\ref{grow}) can exactly reduce to that in \cite{kofman_theory_1997}.  Since the results in \cite{kofman_theory_1997} has been verified to be very accurate when the expansion (\ref{expansion}) is justified by comparing with numerical results, it also provides another check for our results. In addition, the analytical results in \cite{kofman_theory_1997} are only valid for modes inside the broad resonance band, but our results in (\ref{grow}) is a direct conclusion from a more general expression (\ref{particle}) that applies to all the three types of resonances.

\subsection{Narrow resonance}

Another interesting case is the narrow resonance which corresponds to $A_k \gg 2q$ and $e^{\pi \zeta_0^2} \ll 1$. For this case, $\mathfrak{B}$ approximately lies in a very narrow region
\bqn\lb{narrow_condition}
\left(n+\frac{1}{2}\right)\pi - e^{\pi \zeta_0^2/2} <\mathfrak{B} < \left(n+\frac{1}{2}\right)\pi + e^{\pi \zeta_0^2/2},\;
\eqn
and its width can be approximately expressed as $\Delta \mathfrak{B} \simeq 2 e^{\pi \zeta_0^2/2}$. From (\ref{narrow_condition}), by dropping terms with $e^{\pi \zeta_0^2/2}$ we observe that the resonance bands approximately locate at $\mathfrak{B} \simeq n \pi + \frac{1}{2} \pi$, which gives $A_k \simeq n^2$. This implies the resonance only happens at the narrow ranges of $A_k$ and $q$ around $ A_k \simeq n^2$. The particle production rate and field amplification for this case are given by
\bqn
n_k^{(j+1)} &=& \frac{1}{4}\left(\sqrt{1+e^{\pi \zeta_0^2}} + e^{\pi \zeta_0^2/2}\right)^{2j+2},\\
A_{j+1} &\simeq& \frac{1}{2} \left(\sqrt{1+e^{\pi \zeta_0^2}} + e^{\pi \zeta_0^2/2}\right)^{2j+2}.\lb{Aj1_narrow}
\eqn
Although $e^{\pi \zeta_0^2/2} \ll 1$, the particle production and field amplification can still be enhanced if there are a large number of oscillations (c.f. the bottom panel of Fig.~\ref{fit}). For a fixed value of $q$, $\zeta_0^2$ is monotoning decreasing with respect to $A_k$. This indicates that for the same number of oscillation, the modes in the first band ($n=1$) is more significant than others. 

\section{Comments on parametric resonance during inflation and reheating}
The formalism developed in the above can be applied to a lot of scenarios that are related to the inflationary cosmology. Here we consider its potential applications to three examples: the sound resonance during inflation \cite{cai_primordial_2018b}, the particle production during reheating \cite{greene_structure_1997, kofman_theory_1997}, and the generations of oscillons for a self-resonance inflaton field \cite{liu_gravitational_2018, antusch_oscillons_2018}.


The sound resonance is studied recently in \cite{cai_primordial_2018b}, which leads to a novel resonance mechanism for generations of large curvature perturbations $\mathcal{R}_k$ during inflation. In general, the evolution of the Fourier mode $u_k=z \mathcal{R}_k$ associated with the curvature perturbation $\mathcal{R}_k$ during the slow-roll inflation obeys the following perturbation equation,
\bqn
\frac{d^2u_k}{d \tau^2}+ \left(c_s^2 k^2 - \frac{z''}{z} \right)u_k=0,
\eqn
where $z= \sqrt{2 \epsilon_1} a/c_s$ with $\epsilon_1 = - \dot H/H^2$, $\tau$ denotes the conformal time, and $c_s$ represents the nontrivial sound speed of the perturbation mode. In the sound resonance model proposed in \cite{cai_primordial_2018b}, the authors consider a phenomenological parametrization of $c_s^2$ in the form $c_s^2 = 1- 2 \xi [1-\cos{(2 k_* \tau)}]$ with $\xi$ is the amplitude of the oscillation and $k_*$ is the oscillation frequency. Considering that the amplitude $\xi$ of the oscillation is assumed to be very small, then the equation of motion for the propogating mode $u_k=z \mathcal{R}_k$ on sub-Horizon scales can be casted into the Mathieu equation (\ref{mathieu}) \cite{cai_primordial_2018b},
\bqn
\frac{d^2u_k}{dx^2} + (A_k - 2q \cos{2 x})u_k=0\nb
\eqn
with $A_k = k^2/k_*^2+2 q - 4 \xi$, $q = 2 \xi -\xi  k^2/k_*^2$, and $x = k_* \tau$. Applying the formulas of the amplification, we find that the primordial perturbation spectrum can be estimated by
\bqn
P_\mathcal{R}(k) \simeq A_{j+1}\times A_s \left(\frac{k}{k_p}\right)^{n_s-1},
\eqn
where $A_s = \frac{H^2}{8 \pi^2 M_{\rm Pl}^2 \epsilon}$ is the amplitude of power spectrum as in the standard slow-roll inflation, $n_s$ is the corresponding spectral index at pivot scale $k_p$ and the field amplification $A_{j+1}$ can be interpreted by (\ref{Aj1}) with the number of the oscillations given by $j+1 = \frac{k_*}{\pi k} (e^{\Delta N}-1)$, where $\Delta N$ denotes the number of e-folds from the beginning of oscillation to the horizon crossing for mode $k$. This, on the other hand, provides the second explanation of the significance of the first band $(n=1)$ since it has more oscillations than other bands. 


After inflation, the inflaton $\phi(t)$ becomes oscillating around the minimum of its potential, i.e., $\phi(t) \simeq \bar \phi \sin{(m_\phi t)}$. This oscillating behavior leads to parametric resonance during the early stages of reheating, giving rise to copious particle production in fields coupled to it. For simplicity, we consider the coupling of the inflaton to the scalar 􏹝field $\chi$, through an interaction term of the form, 
\bqn
V(\phi, \chi) = \frac{1}{2} m_\phi \phi^2 + \frac{1}{2}g^2 \phi^2 \chi^2,
\eqn
where $g$ denotes the coupling between the inflaton field $\phi$ and the scalar field $\chi$. Then the equation of motion for the Fourier mode $\chi_k$ obeys the modified Klein-Gorden equation,
\bqn
\frac{d^2\chi_k}{dt^2} + 3 H \frac{d\chi_k}{dt} + \left(\frac{k^2}{a^2}+ g^2 \phi^2(t)\right)\chi_k=0,
\eqn
where $t$ denotes the cosmic time. Defining $u_k=a^{3/2}\chi_k$, this equation can be approximately described by (\ref{mathieu}),
\bqn
\frac{d^2u_k}{dx^2} + \left(A_k - 2q \cos(2 x)\right) u_k=0,\nb
\eqn
with $A_k=2q+k^2/(m_\phi^2 a^2)$, $q=g^2\bar \phi^2/(4 m_\phi^2)$, and $x=m_\phi t$. In deriving this we neglected $-\frac{3}{4}(2 \ddot a/a+\dot a^2/a^2)$, which is not important relative to the $g^2 \phi^2$ term during preheating \cite{bassett_inflation_2006}. The particle production rate $|n_k^{j+1}|^2$ can be analyzed  from (\ref{betaj1}) with the number of oscillations $j+1=m_\phi t /\pi$. The strength of resonance depends on $A_k$ and $q$. Small $q (\lesssim 1)$ corresponds to narrow resonance for which the width of the instability band is very small as shown in Fig.~\ref{widthB}. For large $q (\gg 1)$ the broad resonance can occur for a wide range of the parameter space.  The tachyonic resonance is also allowed during reheating process if one replaces the interaction 􏸟$(1/2)g^2 \phi^2 \chi^2$ by 􏸟$(1/2)g \phi^2 \chi^2$ and allows $g<0$. When $|g|$ is large enough to make $A_k <2q$, then the tachyonic resonance occurs for which both the particle production rate and field amplification can be dramatically enhanced \cite{greene_inflaton_1997, Kofman:2001rb, Abolhasani:2009nb, Dufaux:2006ee}. 

The oscillating inflaton can also become self-resonance if it has a self-resonance potential, with which the inflaton field perturbations $\delta \phi_k$ obeys the equation of motion, 
\bqn
\delta \ddot \phi_k + (k^2/a^2+V''(\phi))\delta \phi_k=0,
\eqn
and can be amplified as the inflaton oscillates about the minimum of its potential. Since $V''(\phi)$ is periodical for these self-resonance potentials, they provide a similar parametric resonance mechanism for amplifying $\delta \phi_k$ at certain momentum space and generating the oscillons. When the perturbation modes are initially displaced in the tachyonic region ($k^2/a^2+V''(\phi)<0$) and then enter periodically into this region, these modes can be  dramatically amplified due to the tachyonic resonance. These process have also been known as {\em tachyonic preheating} and {\em tachyonic oscillations} as mentioned in \cite{antusch_oscillons_2018}. For the modes with $k^2/a^2+V''(\phi)>0$ always satisfied, they belong to the parametric resonance with $A_k >2q$ and the resonating amplification can only occur at certain ranges of frequencies that satisfy the resonance condition (\ref{consition}). Such amplifications have gained much attentions recently since it could provide interesting mechanisms for generating large GWs that could be detected by aLIGO-Virgo networks \cite{liu_gravitational_2018, fu_production_2018}.

Note that we have treated $A_k$ and $q$ as constants. In fact, they are decaying with the expansion of the Universe. In this case, the field modes in each oscillation can still be given by (\ref{j_sol}) but with $\zeta_0 \to \zeta_0^{(j)}$ being different in each oscillation. Similarly one needs to make substitute in each oscillation as $\mathfrak{B} \to \mathfrak{B}_j^{j+1}$, $Y_{\pm} \to Y_{\pm}^{(j)}$, $Z_{\pm} \to Z_{\pm}^{(j)}$, and $\kappa \to \kappa_j$. With such extension, our analysis developed in this paper can analytically trace the whole evolutions of the corresponding field modes $u_k$ starting from the tachyonic resonance (with large amplitude of inflaton field initially i.e. large $q$) until the narrow resonance (small $q \ll 1$). The decaying of the oscillations also brings complications in determining the resonance condition. This is because the modes may only grow in some of oscillations, keeping non-growing in others. Therefore, among all of oscillations, the essential question now is to determine how many oscillations that the relevant modes can be amplified. We would like to address this issues in details in our future works.

\section{Summary and Outlook}
To summarize, we have presented a quantitative and general analysis of parametric resonance of the relevant field modes evolving during inflation and reheating. This analysis gives a condition for the occurrence of the resonance and provides clear and quantitative explanation for the  field amplification and particle productions. Further issues including the high-order approximations, effects of decaying oscillations, back reaction of quantum fluctuations on the background spacetime metric, observational predictions of the resonating field modes can also be studied based on our analysis.  Our result is general and simple to use, and has applications beyond the inflation related context, for example, to the production of vector dark matter \cite{dror_parametric_2018, co_qcd_2018, co_dark_2018} by parametric resonance, to the production of electron-positron pairs from vacuum by a periodical laser pulses \cite{dumlu_stokes_2010, hebenstreit_momentum_2009, fillion-gourdeau_resonantly_2013}, and to nonequilibrium quantum field theory \cite{calzetta_nonequilibrium_2008, berges_parametric_2003}.

\section*{Acknowledgements}
The authors thank Drs. Qing-Guo Huang, Jing Liu, and Bao-Fei Li  for useful discussions. This work is supported by National Natural Science Foundation of China with the Grants Nos. 11675143 (T.Z. \& Q.W.), 11675145 (A.W.), and the Fundamental Research Funds for the Provincial Universities of Zhejiang in China with Grants No. RF-A2019015 (T.Z. \& Q.W.).


%

\end{document}